\documentclass[aps,prd,floatfix,showpacs,nobibnotes,twocolumn,
nobalancelastpage,superscriptaddress]{revtex4}
\usepackage{graphicx}\usepackage{amsmath}
\usepackage{times}
\usepackage[T1]{fontenc}\usepackage{textcomp}\usepackage{mathptmx}

\begin{document}


\title{Guaranteed and Prospective Galactic TeV Neutrino Sources}

\author{Matthew D. Kistler}
\email{kistler@mps.ohio-state.edu}
\affiliation{Department of Physics, The Ohio State University, Columbus, Ohio
43210}

\author{John F. Beacom}
\email{beacom@mps.ohio-state.edu}
\affiliation{Department of Physics, The Ohio State University, Columbus, Ohio
43210}
\affiliation{Department of Astronomy, The Ohio State University, Columbus, Ohio
43210}

\date{September 18, 2006}
\smallskip

\begin{abstract}
Recent observations, particularly from the HESS Collaboration, have revealed
rich Galactic populations of TeV gamma-ray sources, including a collection
unseen in other wavelengths.  Many of these gamma-ray spectra are well measured
up to $\sim 10$~TeV, where low statistics make observations by air \v{C}erenkov
telescopes difficult.  To understand these mysterious sources, especially at
much higher energies--where a cutoff should eventually appear--new techniques
are needed.  We point out the following: (1) For a number of sources, it is very
likely that pions, and hence TeV neutrinos, are produced; (2) As a general
point, neutrinos should be a better probe of the highest energies than gamma
rays, due to increasing detector efficiency; and (3) For several specific
sources, the detection prospects for km$^3$ neutrino telescopes are very good,
$\sim$~1--10 events/year, with low atmospheric neutrino background rates above
reasonable energy thresholds.  Such signal rates, as small as they may seem,
will allow neutrino telescopes to powerfully discriminate between models for the
Galactic TeV sources, with important consequences for our understanding of
cosmic-ray production.
\end{abstract}



\maketitle

\section{Introduction} 
The field of TeV gamma-ray astronomy is exploring energy regimes that have been,
until recently, out of reach to astrophysicists.  Yet, even as the catalog of
TeV sources continues to grow, it is still debated whether the observed gamma
rays are produced leptonically, through the inverse Compton scattering of
energetic electrons on ambient photons ($e^-\gamma~\rightarrow~\gamma\,e^-$), or
hadronically, though neutral pion decay ($\pi^0 \rightarrow \gamma\gamma$).  Air
\v{C}erenkov telescopes, such as HESS~\cite{Hinton:2004eu}, can measure a source
spectrum with high precision for $E_\gamma\sim$ $1-10$~TeV.  One might hope that
as we probe higher energy gamma rays, indications of their true origin would be
revealed.  However, at energies $\gtrsim 10$~TeV, the difficulties of gamma-ray
astronomy become more pronounced, due to the low statistics of the quickly
declining signal spectra.

It is well established that a distinctive feature of a \textit{pionic}
(hadronically-produced) gamma-ray spectrum is an accompanying flux of
neutrinos~\cite{Stecker:1978ah, Gaisser:1994yf, Learned:2000sw, Halzen:2002pg}. 
These neutrinos originate from the decay of charged pions ($\pi^+$, $\pi^-$),
which are produced in approximately equal numbers with neutral pions in
proton-proton scattering.  Relative to gamma-ray telescopes, the new km$^3$
neutrino telescopes~\cite{Ahrens:2002dv, Katz:2006wv} will have several
advantages that result in improved performance at the highest energies.  The
rapidly falling atmospheric neutrino background, rising neutrino-nucleon cross
section ($\sigma_{\nu N}\sim E_\nu$), and increasing muon range, which
effectively expands the (already large) detector volume ($R_\mu \sim
\ln{E_\mu}$), all help to amplify the diminished flux at these energies.  In
fact, as the background quickly becomes negligible in the TeV range, the
detection of \textit{any} high energy neutrinos from a source could
significantly indicate a hadronic production mechanism.

\begin{figure}[b!]
\includegraphics[width=3.25in,clip=true]{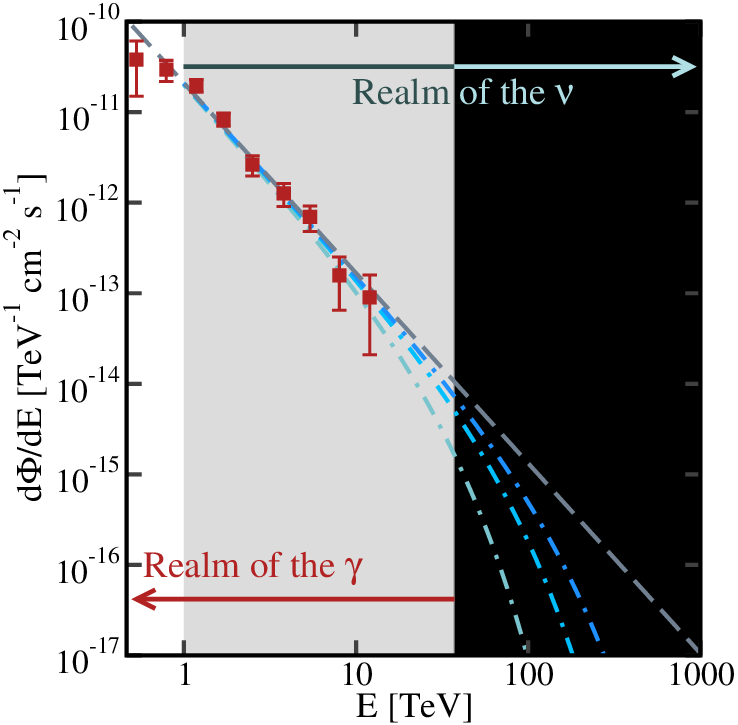}
\caption{\label{rxjdraw} Ranges over which gamma-ray and neutrino observations
can measure a source spectrum ($E_\gamma \simeq E_\pi / 2$, $E_\nu \simeq E_\pi
/ 4$).  Squares are HESS measurements of Vela~Jr., fit by a power law (dashed
line).  Neutrino telescopes can probe higher energies to distinguish between
pionic spectra with cutoffs (dot-dashed lines) by measuring the
$\nu_\mu$-induced muon and shower spectra.  Muon rates of $N(E_\mu> 1$~TeV)
$\sim$ $3-6$ yr$^{-1}$ correspond to the dot-dashed lines (see
Section~\ref{VelaSection}), illustrating the sensitivity of neutrino
observations.}
\end{figure}
%
Neutrino telescopes have capabilities far beyond breaking the degeneracy between
leptonic and hadronic production models.  Spectral features in the highest
energy regime, especially an expected cutoff (related to the maximum accelerated
proton energies), shall not remain inaccessible to observation.  Measurement of
the energies of neutrino-induced muons and showers, which are related to the
original charged pion energy, can probe the source proton spectrum in a
complementary manner to gamma-ray observations, which effectively measure the
neutral pion spectrum~\cite{Lipari:2006uw, AharonianSF06}.  The sensitivities of
these two independent approaches, including the regime where they coincide, are
illustrated in Fig.~\ref{rxjdraw}.  The ability to accurately measure
neutrino-induced muon spectra greatly improves the prospects for detecting point
sources, as the harder source spectra dominate the atmospheric background above
$\sim 1$~TeV.  While muon tracks have better angular resolution ($\lesssim
1^\circ$), neutrino \textit{showers} ($\sim 10^\circ$ in water) more faithfully
trace the spectrum.  Shower observations, which measure the $\nu_e$ and
$\nu_\tau$ fluxes, when combined with muon data, also allow for the study of the
ratio of neutrino flavors arriving from a source~\cite{Beacom:2003nh,
Barenboim:2003jm}.

Considering the latest observations of Galactic TeV sources, we calculate the
corresponding spectra of detectable neutrino-induced muons and showers, assuming
only that the observed gamma-ray spectra are pionic, for a range of possible
high energy cutoffs in the spectra.  Relative to analyses in which the total
numbers of signal and background events are counted (e.g.,
Ref.~\cite{Ackermann:2004ag}), a maximum likelihood analysis would have much
more power.  For example, below $\sim 1$~TeV, a single event has a much greater
probability of being background than at $10$~TeV, where the source signal is
dominant.  Spectra allow for such an approach, which takes full advantage of
experimental data, in studying the high energy behavior of TeV sources. 
Resolving hadronic activity at these extreme energies would provide clear
evidence concerning the sources of Galactic cosmic rays~\cite{Hillas:2005cs}.

Atmospheric cosmic-ray showers give rise to high rates of down-going muons,
forcing a neutrino telescope to search for up-going muons resulting from
neutrino interactions.  IceCube is well-situated to utilize the high resolution
of these $\nu_\mu$-induced muons in observing northern-sky sources.  However, a
detector is needed in the northern hemisphere to accurately locate southern-sky
neutrino sources, although IceCube may also measure shower rates from
particularly bright sources.  Together, IceCube and a km$^3$ Mediterranean
detector will provide full-time coverage of the entire sky, a feature distinct
to neutrino telescopes.  The combined observations from these detectors can be
used to study compound objects, like the Vela complex, by: (1) Discovering
neutrino sources through high-resolution $\nu_\mu$-induced muons; (2) Confirming
agreement with gamma-ray observations in the low energy regime; (3) Examining
previously unexplored energies using muons and showers together.

The sources that we examine in detail are briefly discussed in
Section~\ref{ProspSection}.  Our empirical calculations, based upon observed TeV
spectra, should be compared with previous Galactic neutrino studies, which have
considered theoretical production models or less recent data (e.g.
Refs~\cite{Kolb:1985bb, Crocker:2004nk, Candia:2005nw, Alvarez-Muniz:2002tn,
Costantini:2004ap, Bednarek:2004ky, Aharonian:2005cx, Anchordoqui:2003vc,
Anchordoqui:2005gj, Bertone:2005xz, Bertone:2006nq, Albuquerque:2001jh}).  In
two very recent works, by Lipari~\cite{Lipari:2006uw} and Kappes, Hinton,
Stegmann, and Aharonian~\cite{AharonianSF06}, a similar approach to ours has
been followed, with complementary perspectives and details.
%
%
\section{Promising TeV Neutrino Sources}
\label{ProspSection}
\textbf{\textit{Vela Region}---}Of particular interest amongst prospective
neutrino sources, the shell-type supernova remnant (SNR) Vela~Jr.~(RX
J0852.0--4622) is one of the brightest objects in the southern TeV sky.  The
hard, intense TeV gamma-ray spectrum of Vela~Jr., best explained as being pionic
in nature, makes it an intriguing object to study~\cite{Aharonian:2005sz}. 
Additionally, the morphology of the TeV emission may make this one of the most
interesting astrophysical neutrino sources.  HESS has also measured a TeV
spectrum from Vela~X, the pulsar wind nebula (PWN) associated with the larger
Vela SNR, and advanced a leptonic origin~\cite{Aharonian:2006xx}.  However, if
this spectrum is instead pionic, as proposed in Ref.~\cite{Horns et al.(2006)},
the accompanying neutrino flux would be easily detectable.  The non-detection of
neutrino events from such an intense source would allow for a significant test
of leptonic production in a short period of observation.

\textbf{\textit{Galactic Center Diffuse Emission}---}HESS has also recently
discovered a region of diffuse TeV emission from the Galactic Center
ridge~\cite{Aharonian:2006au}.  This source has several interesting features
which merit further investigation.  The large extent of the emission, hardness
of the spectrum, high gas density (which is well-correlated with the TeV
emission) and strong magnetic fields in the region leave very little doubt that
this spectrum is pionic~\cite{Aharonian:2006au}.  In addition, the total flux
from this region is actually about twice as intense as that of the previously
discovered source coincident with Sgr~A*~\cite{Aharonian:2004wa}.  Measurement
of the accompanying neutrino flux would provide an independent confirmation of
the means of production, something which has never been possible.

\textbf{\textit{Unknown Knowns and Known Unknowns}---}We also examine the
neutrino detection prospects of other known TeV sources, including SNR RX
J1713--3946, the spectrum of which has been measured twice by HESS and
determined to likely be pionic~\cite{Aharonian:2004vr, Aharonian:2005qm}.  The
calculated $\nu_\mu$-induced muon fluxes based upon the two different sets of
HESS data illustrate the effects of different high energy spectral assumptions. 
Recently, HESS has reported the discovery of four TeV sources in the Galactic
Plane which have no apparent counterparts at other
wavelengths~\cite{Aharonian:2005kn, Aharonian:2005rv}.  In addition to these,
EGRET previously discovered numerous unidentified sources of GeV gamma rays in
the Milky Way~\cite{Gehrels:2000gp}.  The confirmed observation of neutrinos
from these sources would also reveal an intimate connection between TeV gamma
rays and cosmic rays.  Furthermore, the understanding of these and other
gamma-ray sources may account for the TeV excess~\cite{Prodanovic:2006bq}.
%
%
\section{Basics of Neutrino Detection}
\label{DetSection}
In high energy $p$-$p$ scattering, $\pi^+$, $\pi^-$, and $\pi^0$ are produced in
nearly equal numbers~\cite{Gaisser:1990vg}.  Gamma rays are the result of the
decay $\pi^0 \rightarrow \gamma\gamma$, while neutrinos originate from the
$\pi^+\rightarrow \mu^+ \nu_\mu\rightarrow e^+ \bar{\nu}_\mu \nu_e \nu_\mu$ and
$\pi^-\rightarrow \mu^- \bar{\nu}_\mu \rightarrow e^- \nu_\mu \bar{\nu}_e
\bar{\nu}_\mu$ decay channels.  The ratio of neutrinos to photons from pion
decay is easily found.  From charged pion decay, the resulting initial neutrino
flavor ratio, $\nu_e:\nu_\mu:\nu_\tau$, is $1:2:0$.  During the traversal of
astrophysical distances, vacuum neutrino oscillations transform this ratio to
$1:1:1$.  In neutrino telescopes, neutrinos and antineutrinos are practically
indistinguishable.  We can then consider their sum, $\nu + \bar{\nu}$, and
average cross section in all calculations.  All further references to neutrinos
will imply $\nu + \bar{\nu}$.  Thus, for equal pion multiplicities, each photon
from $\pi^0$ decay corresponds to one neutrino of each flavor
($N_\gamma=N_{\nu_e}=N_{\nu_\mu}=N_{\nu_\tau}=N_\nu$).  The typical energy of
the neutrinos resulting from these decays is $\sim1/2$ of the gamma-ray energy
from $\pi^0$ decay.  The resulting $\nu + \bar{\nu}$ spectrum is then shifted,
relative to the source gamma-ray spectrum of $d\Phi_\gamma/dE_\gamma =
\phi_\gamma\, E_\gamma^{-\Gamma}$, as
\begin{equation}
  \frac{d\Phi_\nu}{dE_\nu} = \left(\frac{1}{2}\right)^{\Gamma-1} 
  \phi_\gamma\,E_\nu^{-\Gamma} = \phi_\nu\,E_\nu^{-\Gamma}\,,
\end{equation}
where we consider each neutrino flavor separately.

We will generally refer to the normalization of the gamma-ray spectrum as
$\phi_\gamma$, the differential photon flux at $1$~TeV (in TeV$^{-1}$ cm$^{-2}$
s$^{-1}$), which is commonly used in gamma-ray astronomy.  Typical source
gamma-ray spectra have $\phi_\gamma \sim$ $5-20$ $\times\, 10^{-12}$ TeV$^{-1}$
cm$^{-2}$ s$^{-1}$ and $\Gamma \sim$ $1.8-2.4$.  For greater detail on the
relationship between the initial proton spectrum and the resulting product
spectra, see Ref.~\cite{Kelner:2006tc}.  Pions can also be produced, in a
different ratio, in $p$-$\gamma$ scattering, but this is generally only
important at much higher energies~\cite{Anchordoqui:2004eb}.  A neutrino flux
may be observed through neutrino-induced muons and electromagnetic (and
hadronic) cascades, which are referred to as showers.  We shall consider both
methods in further detail.
%
\subsection{Muon Detection}
Our analysis of muon detection will be limited to $\nu_\mu$ charged-current (CC)
events with an observable final energy of $E_\mu > 0.5$~TeV, which can be
detected through their \v{C}erenkov emission with an angular resolution of
$\lesssim 1^\circ$~\cite{Ahrens:2003ix}.  It is important to recognize that the
relevant quantity in these events is the measured energy of the \textit{muon},
which can be reconstructed from radiative losses in the
detector~\cite{Kokoulin:1988mv, Lipari:1991ut, Chikkatur:1997si,
Anchordoqui:2005is}.  As muons can be produced far outside of the detector, with
energy loss prior to entering, results of calculations are given in terms of
$E_\mu$ (at the detector) instead of $E_\nu$.

We can consider $\nu_\mu$-induced muons produced both inside and outside of the
instrumented volume of the detector.  \textit{Contained} event rates can be
found by combining the neutrino flux, detector mass, and $\sigma_{\rm CC}$ for
neutrino-nucleon scattering.  We use the average of the $\nu$-$N$ and
$\bar{\nu}$-$N$ cross sections, as computed in Ref.~\cite{Gandhi:1998ri}, and
$\left\langle y(E_\nu)\right\rangle$ from Ref.~\cite{Gandhi:1995tf}.  We
consider a detector that is entirely composed of, and enclosed by, water (ice),
with a km$^{2}$ detector area, which is a reasonable approximation at these
energies.  Using an IceCube-like effective detector area, such as computed in
Ref.~\cite{Gonzalez-Garcia:2005xw}, would lead to the lower energy muon spectrum
being slightly suppressed.  An actual km$^3$ detector will be situated on top of
solid rock, which will enhance the upward-going neutrino-induced muon rates
relative to our calculations.  Also, muons produced via $\nu_\tau$ CC
interactions, which contribute to the total muon flux through the $\tau^\pm
\rightarrow \mu^\pm \nu_\mu \nu_\tau$ decay channel, will not be
considered~\cite{Gaisser:1990vg, Dutta:2000hh, Beacom:2001xn}.  Our formulas
follow those of Gaisser~\cite{Gaisser:1990vg, Gaisser:1985cm}, with appropriate
approximations.  Considering the uncertainties in both the gamma-ray spectra and
neutrino telescope performance, which may not be fully understood until the
actual detectors are calibrated, as well as the low statistics, our calculations
are sufficient for the scope of this work.

Using these parameters, the spectrum of contained $\nu_\mu$-induced muon events
is calculated as
\begin{equation}\label{eq:cont}
  \left(\frac{dN_\mu}{dE_\mu}\right)_{\rm cont} = 
  \kappa\,V_{\rm det}\,\frac{d\Phi_\nu}{dE_\nu}\, e^{-E_\nu /E_\nu^{\rm cut}} 
  \sigma_{\rm CC}(E_\nu)\,  e^{-\tau_\oplus}\,,
\end{equation}
where $V_{\rm det}$ is the detector volume, $E_\nu^{\rm cut}$ is an assumed
exponential cutoff in the neutrino spectrum, and the term
$\kappa=N_A\,\rho\,T\,\left\langle 1 - y(E_\nu)\right\rangle^{-1}$ takes into
account observation time (T), normalization of the muon spectrum, and the molar
density of water.  The energy of the produced muon is related to the original
neutrino energy as $E_\mu = \left\langle 1 - y(E_\nu)\right\rangle E_\nu$.  The
term $e^{-\tau_\oplus}$, with $\tau_\oplus=N_A\, \lambda_\oplus\,\sigma_{\rm
tot}(E_\nu)$, accounts for neutrino attenuation due to scattering within Earth. 
Here, $\sigma_{\rm tot}=\sigma_{\rm CC} + \sigma_{\rm NC}$ and $\lambda_\oplus$
is the average column depth (in cm.w.e.) of Earth based upon the declination of
the source~\cite{Gandhi:1995tf}.  This attenuation factor only becomes important
at $E_\nu\gtrsim10$~TeV, and varies with declination, as discussed in
Ref.~\cite{Lipari:2006uw}.

A high energy muon born outside of the detector may still be detectable when it
enters the instrumented detector volume.  To find the propagation range of a
muon, we assume an average continuous muon energy loss of
\begin{equation}
	\frac{dE}{dX}=-\alpha-\beta E\,,
\end{equation}
where $\alpha=2.0\times10^{-6}$ TeV cm$^2$ g$^{-1}$ and $\beta=4.2\times10^{-6}$
cm$^2$ g$^{-1}$~\cite{Lipari:1991ut, Dutta:2000hh}.  Integrating the average
energy loss results in a range for a muon of initial energy $E_\mu$ of
\begin{equation}
	R_\mu(E_\mu,E_\mu^f)=\frac{1}{\beta}
	\ln\left[\frac{\alpha+\beta E_\mu}{\alpha+\beta E_\mu^f}\right]\,,
\end{equation}
where $E_\mu^f$ is the energy of the muon as it enters the detector.  As the
muon range (typically a few km) increases with energy, so does the effective
volume of the detector.  The observed through-going spectrum, accounting for the
probability of observing a muon entering the detector with energy $E_\mu$, can
then be calculated as
\begin{eqnarray}
	\left(\frac{dN_\mu}{dE_\mu}\right)_{\rm thru} \!=&\!&
	\!\frac{N_A\,\rho\,T\,A_{\rm det}}{\alpha+\beta E_\mu} \nonumber \\
	&\!& \!\times \int_{E_\mu}^\infty \!dE_\nu
\frac{d\Phi_\nu}{dE_\nu}\,
	e^{-E_\nu /E_\nu^{\rm cut}}\sigma_{\rm CC}(E_\nu)\,
	e^{-\tau_\oplus},
\end{eqnarray}
where $A_{\rm det}$ is assumed to be $1$ km$^2$.  It should be noted that our
use of the total $\sigma_{\rm CC}$ and $\left\langle y \right\rangle$ may
overestimate the muon signal rates by $\sim$~$20-30$\% (perhaps more for spectra
harder than $E^{-2}$).  However, the atmospheric background will also be
overestimated by a similar, even slightly larger amount.
%
\subsection{Shower Detection}
For bright point sources or low surface brightness extended sources, shower
events caused by $\nu_e$, $\nu_\tau$ CC interactions can be used to effectively
determine the source spectrum.  To detect shower events, they must be at least
partially enclosed within the instrumented volume.  The shower signal is
dominated by $\nu_e$ and $\nu_\tau$ CC events, which transfer $\sim 100\%$ of
the original neutrino energy to the shower ($E_{\rm sh}\approx E_\nu$).  The
measurable shower spectrum, given in terms of the observed \textit{shower}
energy, $E_{\rm sh}$,
\begin{equation}\label{eq:showeq}
  \left(\frac{dN_{\rm sh}}{dE_{\rm sh}}\right)_{\rm CC} \!= 
  \! 2 N_A\, \rho\, T\, V_{\rm det}\, \frac{d\Phi_\nu}{dE_\nu}\,
  e^{-E_\nu /E_\nu^{\rm cut}} \sigma_{\rm CC}(E_\nu) e^{-\tau_\oplus},
\end{equation}
then effectively traces the original neutrino flux, as the neutral-current (NC)
events from all three flavors, which only contain $E_{\rm sh} = \left\langle
y_{\rm NC}\right\rangle E_\nu$ (in addition to the smaller $\sigma_{\rm NC}$),
only account for $< 10$\% of the total signal.

Combined shower and muon observations would determine the $\nu_\mu /
(\nu_e+\nu_\tau)$ ratio arriving from a source.  This measured flavor ratio has
many applications, as detailed in Refs.~\cite{Beacom:2003nh, Barenboim:2003jm,
Serpico:2005sz, Winter:2006ce}.  The atmospheric $\nu_e$ flux is only about
$\sim 1/20$ of the $\nu_\mu$ background, which itself only adds to the shower
background through the weaker NC channel, since $\nu_\mu$ CC interactions are
identifiable by the resulting muon tracks.  We will not consider the prompt
background, which is the only source of atmospheric $\nu_{\tau}$, as it only
becomes important at very high energies~\cite{Beacom:2004jb}.  These facts help
offset the lower angular resolution of these events ($\lesssim 10^\circ$ is
expected for a Mediterranean detector~\cite{Hartmann:2006tv}, $\lesssim
25^\circ$ for IceCube~\cite{Ahrens:2003ix}), which sets the background event
rate.  The improved shower resolution of a Mediterranean detector, as compared
to IceCube (which is naturally limited by light scattering in ice), increases
the value of showers in determining the high energy properties of a source.
%
\subsection{Neutrino Spectroscopy}
%
\begin{figure}[b]
\includegraphics[width=3.25in,clip=true]{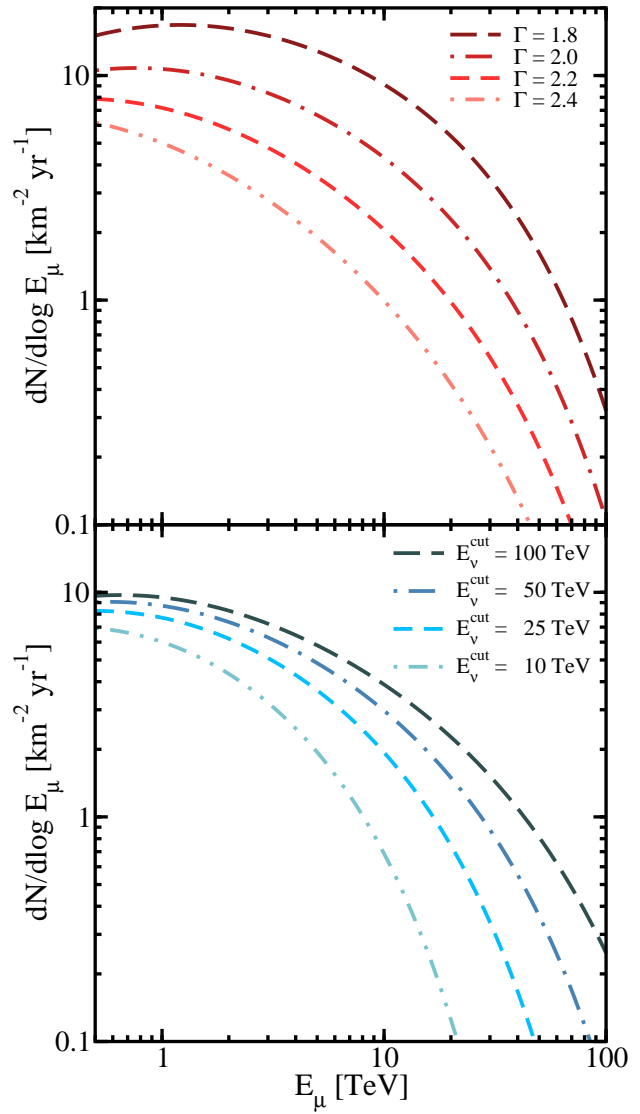}
\caption{\label{gen} A demonstration of the sensitivity of neutrino telescopes
to varying spectral characteristics.  Shown is the (base-10 log) differential
($\nu_\mu + \bar{\nu}_\mu$)-induced muon spectra for a pionic source spectrum
with $\phi_\gamma=20\times10^{-12}$ TeV$^{-1}$ cm$^{-2}$ s$^{-1}$.  In the upper
panel, the lines of decreasing height correspond to spectral indices of
$\Gamma=$ $1.8$, $2.0$, $2.2,$ and $2.4$, with an exponential cutoff $E_\nu^{\rm
cut}=50$~TeV.  In the lower panel, the lines of decreasing height correspond to
neutrino spectrum cutoffs of $100$, $50$, $25$, and $10$~TeV, with a fixed
spectral index of $\Gamma=2.1$.  Each line corresponds to one year of
observation.}
\end{figure}
%
To demonstrate the ability of a neutrino telescope to differentiate between
various spectral properties, we first consider a general object with a fixed
$\phi_\gamma=20\times10^{-12}$ TeV$^{-1}$ cm$^{-2}$ s$^{-1}$ (at $1$~TeV) and
assume a source declination of $\delta = +10^\circ$.  The resulting up-going
muon fluxes for a range of spectral indices and high energy exponential cutoffs,
for one year of IceCube observation, are shown in Fig.~\ref{gen}.  In the top
panel, we consider spectral indices with $\Gamma=$ $1.8$, $2.0$, $2.2$, and
$2.4$, with $E_\nu^{\rm cut}=50$~TeV.  Note that the differences are not
entirely due to differing integrated neutrino fluxes, as the rising $\sigma_{\nu
N}$ results in higher event rates for harder spectra.  In
Ref.~\cite{Lipari:2006uw}, similar results were obtained by normalizing to the
integrated flux above $1$~TeV.  The bottom panel illustrates the effect of
$100$, $50$, $25$, and $10$~TeV neutrino spectrum exponential cutoffs for a
fixed index ($\Gamma=2.1$).  IceCube is well-suited to isolate a similar point
source located in the northern sky through $\nu_\mu$-induced muons within a
short period of operation.  Such a source would also be observable through
shower events by a km$^3$ Mediterranean detector.  The improved shower detection
capabilities of a Mediterranean detector would allow for a more direct
measurement of the arriving neutrino spectrum, adding greatly to the science
that can be extracted from TeV sources.

All of our calculations are done on an empirical basis using the average
measured parameters of gamma-ray source spectra.  All high energy exponential
cutoffs that we will consider are given in terms of $E_\nu^{\rm
cut}$(=$\,$1/2$\,E_\gamma^{\rm cut}$), to allow for a range of calculations to
be compared with future experimental data.  Angle-averaged FLUKA computations of
atmospheric neutrinos past $E_\nu=1$~TeV are used to determine
backgrounds~\cite{Battistoni:2003ju}.  For ease of comparing the expected signal
to background and assessing total rates, we will mainly provide integrated rates
above a given measured energy.  Because the atmospheric spectrum is very steeply
falling at these energies ($\sim E_\nu^{-3.5}$), it is often advantageous to use
a low energy event cutoff $\gtrsim 1$~TeV when calculating significance.  Our
method can also yield the measured muon and shower spectra directly.  Spectral
information should be used for maximum likelihood analysis of observed neutrino
events in a neutrino telescope, to make the best use of measured energies in
determining detection significance.  This spectral information requires
consideration of the energy resolution of the detector.  In IceCube, the energy
resolution for muons and showers is expected to be $\sim 20$\% and $\sim 10$\%
in the logarithm of the energy, respectively~\cite{Ahrens:2003ix,
Kowalski:2005tz}, which does not significantly affect our conclusions.

Note that we present our results for an assumed muon effective area of $1$
km$^2$, and assumed muon angular resolution of $\lesssim 1^\circ$, both
independent of energy.  These results are close to the results of detector
simulations, especially $\gtrsim$ $1$~TeV, the primary area of
interest~\cite{Ahrens:2003ix, Circella:2005fq}.  Below $1$~TeV, these
assumptions are certainly overly optimistic; however, in that range, the
atmospheric background is dominant, and so accuracy is of less importance.  In
regards to the angular resolution, our assumptions are somewhat too conservative
(in particular, the improved resolution at higher energy will help reject
background).  Part of our intent is to show the likely signals, and to encourage
the experimentalists to optimize their detector designs in order to achieve the
necessary sensitivity to observe Galactic TeV sources.  These calculations are
not meant to replace a more detailed study of neutrino source detectability,
which would include time-dependent source locations, zenith-angle-dependent
atmospheric backgrounds, angle-dependent detector sensitivities, event
reconstruction methodologies, stochastic muon energy losses, source modeling,
etc.  Each prospective source should be subjected to such a full Monte Carlo
simulation by the IceCube and Mediterranean collaborations.
%
%
\section{The Vela Complex}
\label{VelaSection}
\begin{figure}[t!]
\includegraphics[width=3.25in,clip=true]{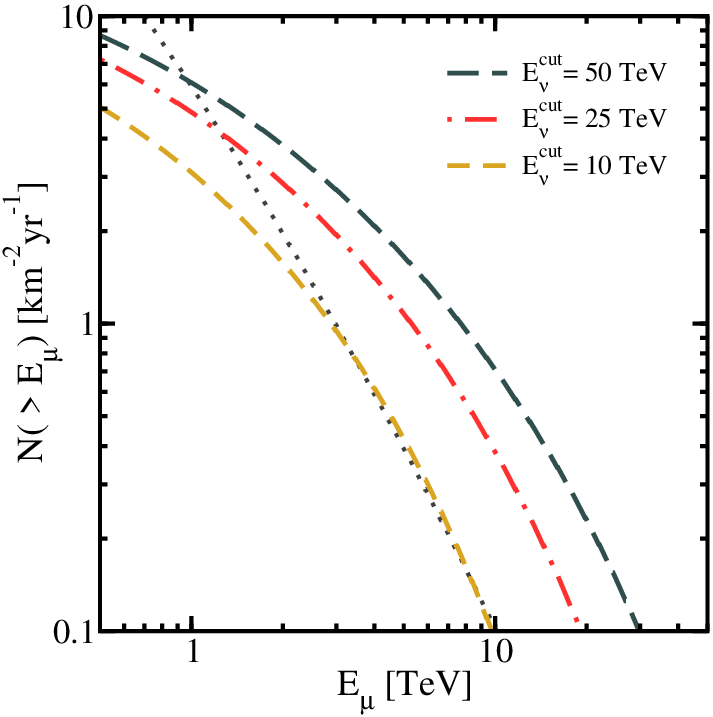}
\caption{\label{rxj0852} \textbf{\textit{Vela~Jr.~(RX J0852.0--4622)---Muons:}}
Integrated ($\nu_\mu + \bar{\nu}_\mu$)-induced muon rates above a given measured
muon energy.  The long dashed (blue), dot-dashed (red), and short dashed (gold)
lines correspond to neutrino spectrum exponential cutoffs of $50$, $25$, and
$10$~TeV, respectively.  The dotted line shows the expected atmospheric
background in a 7 deg$^2$ bin, as discussed in the text.  Rates are for one year
of operation in a km$^3$ Mediterranean detector.}
\end{figure}

Of the many known TeV gamma-ray sources, the shell-type SNR Vela~Jr.~(RX
J0852.0--4622) is one of the most interesting.  This southern-sky source has
been observed at gamma-ray energies exceeding $10$~TeV by HESS.  Their analysis
is suggestive of a hadronic origin for the gamma-ray
spectrum~\cite{Aharonian:2005sz}.  This source is very bright in gamma rays
($d\Phi/dE = 21 \times 10^{-12} (E/$TeV)$^{-2.1}$ TeV$^{-1}$ cm$^{-2}$
s$^{-1}$), with well-defined regions of gamma-ray emission.  Shell-type SNRs are
considered to be the most likely sites of Galactic cosmic-ray proton
acceleration~\cite{Rowell:2005de}.  As these source proton spectra are expected
to be cut off near the knee ($\sim 3 \times 10^{15}$~eV), a cutoff should also
be present for gamma-rays and neutrinos at a lower energy
scale~\cite{Kelner:2006tc}.  Thus, we calculate the expected neutrino-induced
muon rate assuming a pionic spectrum with several neutrino spectrum exponential
cutoffs ($50$, $25$, and $10$~TeV), as shown in Fig.~\ref{rxj0852}.  When
considering emission from the entire source extension, we must accept
atmospheric background from a $\sim 7$ deg$^2$ area.  Comparing the source and
background muon rates, this SNR is expected to be significantly detectable with
a km$^3$ Mediterranean detector.  For the higher cutoffs, discovery may be
possible in only a few years.  These event rates also suggest that ANTARES may
be able to find evidence of neutrino emission from this
source~\cite{Aslanides:1999vq}.

A number of features make Vela~Jr.~unique among prospective TeV neutrino
sources.  The TeV emission is observed to originate from several regions which
are separated by $\sim 2^\circ$.  In addition to detecting the presence of
neutrinos, the intensity of the measured muon spectrum may provide adequate
statistics to allow accurate location of neutrino-production sites.  Thus, this
unique gamma-ray morphology allows for the possible construction of a
\textit{neutrino map} of Vela~Jr.  This capability would enable a comparison of
the production mechanism and high energy activity in different regions of the
same source.  When observing features of the neutrino source, the total
atmospheric neutrino background enclosed is lower, which increases their
individual detectability.  An example of such a map can be seen in
Ref.~\cite{Katz:2006cc}.  Further HESS observations should provide improved
resolution of the TeV emission.

\begin{figure}[t!]
\includegraphics[width=3.25in,clip=true]{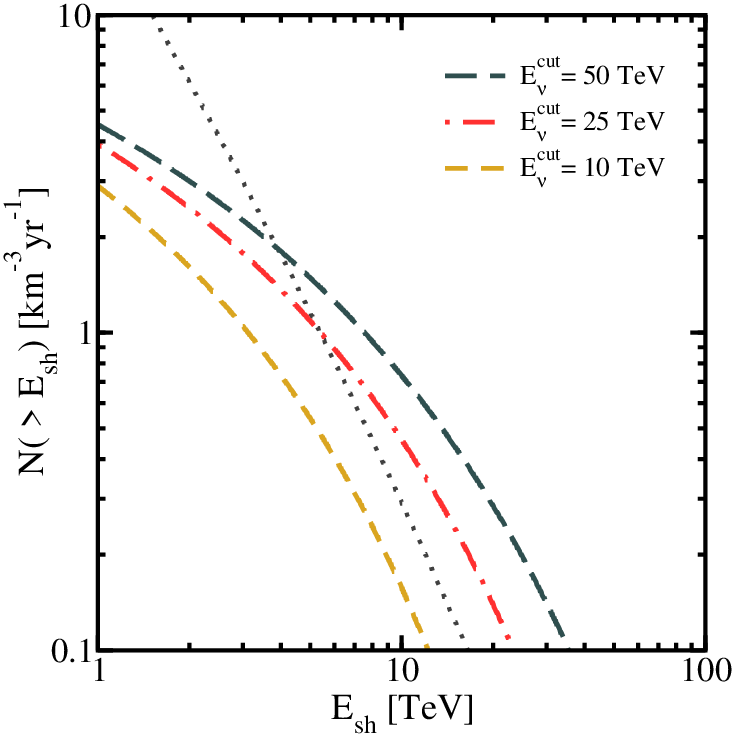}
\caption{\label{rxjshow} \textbf{\textit{Vela~Jr.~(RX
J0852.0--4622)---Showers:}} Integrated neutrino-induced shower rates above a
given measured shower energy, with signal lines as in Fig.~\ref{rxj0852}.  Note
that a different energy scale is used for showers, as measurable shower events
are assumed to have a $1$~TeV energy threshold.  Background rates are for a
circle of $10^\circ$ radius, corresponding to the shower angular resolution in a
km$^3$ Mediterranean detector.  A significant reduction in background results
from improved angular resolution.  Rates are for one year of operation.}
\end{figure}

Additionally, the brightness of the source may also make it possible to observe
appreciable numbers of showers.  As a consequence of the detected showers
directly tracing the neutrino flux, spectral information may be found (even with
a slightly higher background).  Fig.~\ref{rxjshow} shows the rate of showers for
one year of observation, compared to the irreducible atmospheric background. 
With enough observation time, it may be possible to determine the neutrino
flavor ratio from this source by observing the $\nu_\mu$ spectrum with a km$^3$
Mediterranean detector and by also utilizing IceCube to effectively increase the
shower volume (since shower events must be located within the detector) in
measuring the $\nu_e + \nu_\tau$ flux.

%
%
Vela~Jr.~is coincident with a region of the larger Vela SNR.  HESS has also
reported the discovery of TeV gamma rays originating from the Vela PWN, referred
to as Vela~X (of no relation to the HMXB Vela~X-1)~\cite{Aharonian:2006xx}. 
Vela~X has a very hard spectrum, which is best fit with a spectral index of
$\Gamma=1.45$ with an exponential cutoff beginning at $E_\gamma^{\rm
cut}=13.8$~TeV.  In their discovery paper, HESS concludes that this emission is
produced by an inverse Compton mechanism.  An independent analysis has
considered the possibility that this spectrum is produced hadronically, with an
accompanying flux of neutrinos~\cite{Horns et al.(2006)}.  This source is
sufficiently separated from Vela~Jr.~to independently test for neutrino emission
through muons.  If we assume the entire measured flux is pionic, then the
expected $\nu_\mu$-induced muon rate is $N(E_\mu > 1$~TeV) $\sim 4.5$ yr$^{-1}$,
in a km$^3$ Mediterranean detector, which would be relatively easy to measure. 
While our calculation may somewhat overstate the muon rate for such a hard
spectrum, nevertheless, the nondetection of a significant neutrino event excess
would indicate leptonic production.

Considering the large apparent size of the Vela SNR shell ($\sim 8^\circ$), any
TeV gamma-ray emission originating from it may be too diffuse to detect
directly.  Searching for it in neutrinos is a possibility.  A
northern-hemisphere detector would isolate any neutrino point source spectra
located in the region through $\nu_\mu$-induced muons.  Showers have the ability
to observe this entire region simultaneously, raising the intriguing prospect of
utilizing IceCube and a km$^3$ Mediterranean detector in concert to discover and
study bright, but highly extended TeV neutrino sources.
%
%
\section{Galactic Center Region}
The HESS discovery~\cite{Aharonian:2006au} of the region of Galactic Center
diffuse emission (GCD) is important for neutrino, gamma-ray, and cosmic-ray
astrophysics~\cite{Busching:2006qx, Liu:2006bf}.  The spectrum of the GCD was
measured over a very large region spanning the Galactic coordinates
$|l|<0.8^\circ$, $|b|<0.3^\circ$ with
$d\Phi/dE=1.73\times10^{-8}~(E/$TeV)$^{-2.29}$ TeV$^{-1}$ cm$^{-2}$ s$^{-1}$
sr$^{-1}$.  A number of large, dense ($n_H\sim10^4$ cm$^{-3}$) molecular clouds
are known to fill this region~\cite{Tsuboi et al.(1999)}.  Considering the vast
extent of this hard emission ($\sim$~few hundred pc across), along with the high
target density and magnetic fields in the region, the only reasonable production
mechanism for these gamma rays is neutral pion decay.  The close morphological
correlation between the TeV emission and the gas distribution is particularly
compelling~\cite{Aharonian:2006au}.  This TeV gamma-ray spectrum implies a local
cosmic-ray spectrum that is much harder and $3-9$ times denser than that of the
CR flux as measured at Earth.  The likely source of these cosmic-ray protons is
Sgr~A, which hosts the remnant of a recent supernova~\cite{Maeda et al.(2002),
Fryer:2005qn}.  These pioneering conclusions were reached in
Ref.~\cite{Aharonian:2006au}.

In order to calculate the expected $\nu_\mu$-induced muon rate, we must first
take into account the extent of the GCD source.  The muon angular resolution of
a neutrino telescope at these energies ($\sim1^\circ$) covers the entirety of
the GCD.  In effect, this region can be treated as a neutrino point source. 
Integrating over the $\sim1$ deg$^2$ source region, the total photon flux is
$d\Phi/dE=5.2\times10^{-12} (E/$TeV)$^{-2.29}$ TeV$^{-1}$ cm$^{-2}$ s$^{-1}$. 
We define the differential flux at $1$~TeV as $\phi_{\rm GCD}=5.2\times10^{-12}$
TeV$^{-1}$ cm$^{-2}$ s$^{-1}$.  We will not consider showers from the GCD;
however, any such study should also take into account a similar diffuse emission
around Sgr~B~\cite{Aharonian:2006au}.

\begin{figure}[b]
\includegraphics[width=3.25in,clip=true]{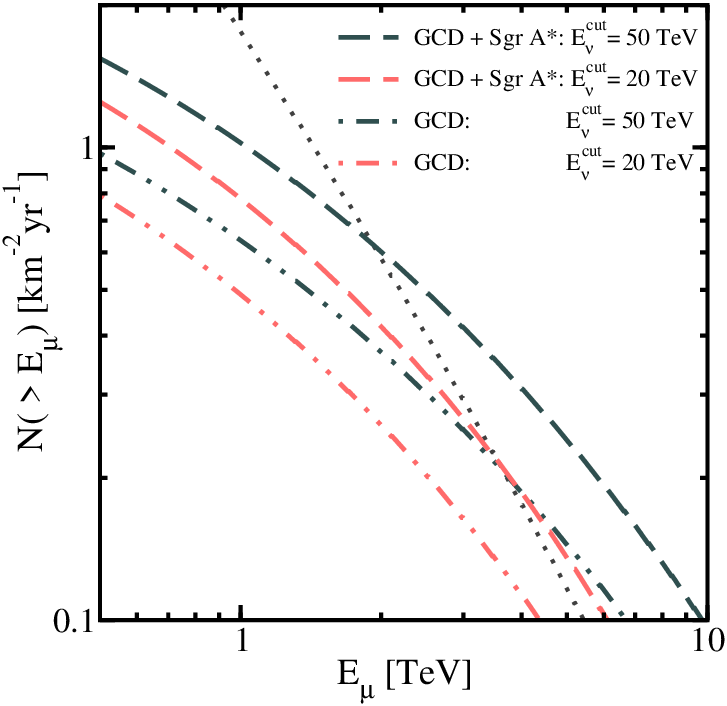}
\caption{\textbf{\textit{Galactic Center region:}} Integrated ($\nu_\mu +
\bar{\nu}_\mu$)-induced muon rates above a given measured muon energy.  Shown
are the rates from the Galactic Center diffuse source (GCD) alone (dot-dashed)
and the result of including the Sgr~A* source (dashed).  Two cases are
considered: a neutrino spectrum exponential cutoff $E_\nu^{\rm cut}=50$~TeV and
a more conservative cutoff of $20$~TeV.  The dotted line shows the estimated
atmospheric background in a 3 deg$^2$ bin.  Rates are for one year of
observation in a km$^3$ Mediterranean detector.}
\label{GCplot}
\end{figure}
%
We first consider the conservative assumption that the GCD spectrum exhibits an
exponential neutrino spectrum cutoff $E_\nu^{\rm cut}=20$~TeV.  As a cosmic-ray
production site, we also consider a higher cutoff of $50$~TeV.  The expected
muon rates are shown in Fig.~\ref{GCplot}.  Even though its angle-integrated
flux is larger, the GCD was revealed only after subtracting the previously
discovered TeV emission from Sgr~A* and a distant SNR~\cite{Aharonian:2004wa,
Aharonian:2005br}.  The TeV gamma-ray flux from Sgr~A*, which has $\phi_\gamma
\sim 0.5~\phi_{GCD}$, can reasonably be assumed to also be pionic, as the
spectral index ($\Gamma=2.2$) is close to the GCD value~\cite{Aharonian:2004wa,
Albert:2005kh}.  As the two sources are coincident, we combine their expected
muon rates, as also shown in Fig.~\ref{GCplot}.  Such an approach to calculating
the total neutrino flux was also taken in Ref.~\cite{Cavasinni:2006nx}, where
the integrated muon rate was found.  The declination of the source, along with
the latitude of the detector, determine the fraction of the time that the source
is below the horizon~\cite{Bertone:2006nq}.  For the Galactic Center this is
$\sim 0.7$, which is taken into account in Fig.~\ref{GCplot}.  The signal rates
shown in Fig.~\ref{GCplot}, when compared to the expected number of atmospheric
events, suggest that this source is near the threshold of discovery for the
lifetime of a km$^3$ Mediterranean detector.  Discovery may be hastened by
utilizing maximum likelihood methods in comparing the expected signal with
background for each muon event with a measured energy, as higher energy events
have a much greater significance.  Such an analysis is only possible with
spectral information.
%
%
\section{A Multitude of TeV Sources}
Numerous discoveries of new gamma-ray sources should be expected from the HESS,
MAGIC and VERITAS air \v{C}erenkov telescopes and the GLAST
satellite~\cite{Baixeras:2003xr, Weekes:2001pd, Gehrels:1999ri}.  However, the
characteristics of the source spectra above $\sim10$~TeV are difficult to
observe.  RX J1713--3946 is also a shell-type SNR, first observed in the TeV
regime by CANGAROO~\cite{Enomoto:2002xk}.  It has since been observed twice by
HESS with greatly improved statistics.  We consider both HESS observations to
illustrate the potential of neutrino telescopes to determine the high energy
behavior of TeV sources.  The flux was initially reported up to $10$~TeV as a
power law ($\Gamma\sim2.19$), with analysis favoring hadronic
production~\cite{Aharonian:2004vr}.  Subsequent observations revealed a spectrum
that extends to at least $40$~TeV with steepening past $\sim10$~TeV that can be
fit by either an energy dependent spectral index or an exponential
cutoff~\cite{Aharonian:2005qm}.  Previous neutrino studies of this source were
inspired first by the CANGAROO detection~\cite{Alvarez-Muniz:2002tn} and later
by utilizing the more accurate 2004 HESS spectrum along with a more detailed
calculation~\cite{Costantini:2004ap}.  We analyze the $\nu_\mu$-induced muon
spectra from these two data sets independently, assuming $E_\nu^{\rm
cut}=50$~TeV for the 2004 data.  As seen in Table~\ref{tab:hess}, the integrated
rates above $1$~TeV are similar.  However, at higher energies calculations based
upon the old power law spectrum exhibit a deviation from the new data.  From the
2004 data, we find $N(E_\mu > 10$~TeV) $\sim 0.3$ yr$^{-1}$, while the 2005 data
yields a rate only $\sim 1/10$ of this, which highlights the sensitivity of
neutrino observations to the spectrum at the highest energies.  Based on either
spectrum, this SNR should be detectable in a km$^3$ Mediterranean detector.

Amongst the HESS catalog of TeV sources are several with no observed counterpart
at any other wavelength.  Explaining the origin of these sources is a
theoretical challenge.  It is possible that they represent a non-trivial new
class of astrophysical objects.  With so little being known about them at
present, the observation of neutrinos from these sources would yield invaluable
insight into their nature.  The first of these sources is HESS
J1303--631~\cite{Aharonian:2005rv}.  Chandra observations revealed no likely
x-ray counterpart~\cite{Mukherjee:2005yp}.  This source is always visible to a
Mediterranean detector.  Three sources were also found in the HESS survey of the
Galactic Plane that remain unidentified: HESS J1614--518, HESS J1702--420, and
HESS J1708--410~\cite{Aharonian:2005kn}.  The spectral information and resulting
integrated $\nu_\mu$-induced muon rates from these sources are given in
Table~\ref{tab:unid}.  It has been proposed that HESS J1303--631 is a remnant of
a Galactic gamma-ray burst (GRB), with a harder ($\Gamma \sim 2.2$) and more
extended emission than has been observed~\cite{Atoyan:2005cc}.  From the given
predictions of the proposed hadronic source model, we would expect $N(E_\mu>
1$~TeV) $\sim 2.1$ yr$^{-1}$, which is potentially verifiable with a km$^3$
Mediterranean detector.  From the measured spectra, detecting these sources
individually may be difficult; however, the stacking method may prove
advantageous.  Stacking effectively increases the total observation rate by
combining the individual event rates from identical sources~\cite{Brandt,
Cillis:2004ks, Brand:2005zf, Achterberg:2005fs}.  Stacking of the observed muon
data may improve the prospects of determining whether neutrinos are present and
help unveil the mystery of these unknown sources.

Other Galactic TeV sources with probable counterparts may also be examined as
potential neutrino sources.  Many new sources were found as a result of the HESS
inner Galactic Plane survey, some of which may be SNRs~\cite{Aharonian:2005kn}. 
MAGIC and VERITAS observations will reveal sources in the outer regions of the
Plane, which has yet to be examined with an instrument of their sensitivity. 
Assuming that the gamma-ray spectra already observed by HESS are pionic, the
calculated neutrino event rates for a number of sources are shown in
Table~\ref{tab:hess}.  Several of these have sufficient rates to consider as
possible neutrino sources independently.  However, it would be difficult to
significantly detect dimmer sources over the atmospheric background.  Stacking,
in combination with a maximum likelihood spectral analysis, may be used to
examine the entire group, or subgroups, of the lower-flux sources.
%
\begin{table}[t]
\caption{\textit{Unidentified HESS Sources:} Integrated ($\nu_\mu +
\bar{\nu}_\mu$)-induced muon rates, assuming a pionic spectrum, for a $1$~TeV
$E_\mu$ threshold and neutrino spectrum exponential cutoff $E_\nu^{cut}$.  For
one year of km$^3$ Mediterranean detector operation, accounting for observable
time below the horizon.  The HESS differential flux at $1$~TeV is given in terms
of $10^{-12}$ TeV$^{-1}$ cm$^{-2}$ s$^{-1}$, with spectral index $\Gamma$.}
\begin{ruledtabular}
	\begin{tabular}{lrlrr}
	 Source & $\phi_\gamma$ & $\Gamma$ & $E_\nu^{cut}$ (TeV) &
$N_\mu(>1$~TeV) \\
 \hline
	 HESS J1303--631 & 4.3 & 2.44 & 10  & 0.3 \\
	                 &     &      & 50  & 0.5 \\
	 HESS J1614--518 & 7.0 & 2.46 & 10  & 0.5 \\
	                 &     &      & 50  & 0.8 \\
	 HESS J1702--420 & 2.5 & 2.31 & 25  & 0.3 \\
	                 &     &      & 50  & 0.4 \\
	 HESS J1708--410 & 1.5 & 2.34 & 10  & 0.1 \\
	                 &     &      & 50  & 0.2 \\
	\end{tabular}
\end{ruledtabular}\label{tab:unid}
\end{table}
%
\begin{table}[t]
\caption{\textit{HESS Sources With Probable Counterparts:} Integrated ($\nu_\mu
+ \bar{\nu}_\mu$)-induced muon rates, assuming a pionic spectrum, for a $1$~TeV
$E_\mu$ threshold and neutrino spectrum exponential cutoff $E_\nu^{cut}$.  For
one year of km$^3$ Mediterranean detector operation (unless noted), accounting
for observable time below the horizon.  The HESS differential flux at $1$~TeV is
given in terms of $10^{-12}$ TeV$^{-1}$ cm$^{-2}$ s$^{-1}$, with spectral index
$\Gamma$.}
\begin{ruledtabular}
	\begin{tabular}{lrlrr}
	 Source & $\phi_\gamma$ & $\Gamma$ & $E_\nu^{cut}$ (TeV) &
$N_\mu(>1$~TeV) \\
\hline
	 Vela Jr.          & 21.0& 2.1  & 10  & 3.1 \\
	 (RX J0852.0--4622)&     &      & 25  & 4.9 \\
	                   &     &      & 50  & 6.1 \\ 
   GC Diffuse        &  5.2& 2.29 & 20  & 0.5 \\
	                   &     &      & 50  & 0.7 \\
	 (+ GC Source)     &     &      & 20  & 0.8 \\
	                   &     &      & 50  & 1.0 \\     
\hline
	 RX J1713.7--3946  & 15.0& 2.19 & 50  & 2.8 \\
	                   & 20.4& 1.98 & 6   & 2.2 \\
	 Vela X            & 9.0 & 1.45 & 7   & 4.5 \\
	 Crab (IceCube)    & 33.0& 2.57 & 50  & 2.7 \\
	 HESS J1514--591   & 5.7 & 2.27 & 25  & 0.9 \\
	                   &     &      & 50  & 1.1 \\
	 HESS J1616--508   & 6.0 & 2.35 & 10  & 0.5 \\
	                   &     &      & 50  & 0.9 \\
	 HESS J1632--478   & 5.5 & 2.12 & 10  & 0.8 \\
	                   &     &      & 50  & 1.5 \\
	 HESS J1634--472   & 2.0 & 2.38 & 10  & 0.2 \\
	                   &     &      & 50  & 0.3 \\
	 HESS J1640--465   & 3.0 & 2.42 & 10  & 0.2 \\
	                   &     &      & 50  & 0.4 \\
	 HESS J1745--303   & 2.5 & 1.8  & 10  & 0.5 \\
	                   &     &      & 50  & 1.2 \\
	 HESS J1804--216   & 4.7 & 2.72 & 10  & 0.1 \\
	                   &     &      & 50  & 0.2 \\
	 HESS J1813--178   & 2.7 & 2.09 & 10  & 0.3 \\
	                   &     &      & 50  & 0.5 \\
	 HESS J1825--137   & 6.0 & 2.46 & 25  & 0.4 \\
	                   &     &      & 50  & 0.5 \\
	 HESS J1834--087   & 2.5 & 2.45 & 10  & 0.1 \\
	                   &     &      & 50  & 0.2 \\
	 HESS J1837--069   & 5.0 & 2.27 & 10  & 0.3 \\
	                   &     &      & 50  & 0.6 \\
	\end{tabular}
\end{ruledtabular}\label{tab:hess}
\end{table}
%
%
\section{A TeV Neutrino Beacon}
It would be beneficial to have a bright source of TeV neutrinos that can be seen
by both IceCube and a Mediterranean detector.  The ideal source for this purpose
would have a declination in the range $+10^\circ \lesssim \delta \lesssim
+30^\circ$, which also has the benefit of reduced neutrino attenuation in Earth
at high energies, resulting in increased muon and shower event rates.  The Crab
nebula is already an accepted standard candle in TeV gamma-ray astronomy.  From
HESS measurements~\cite{Masterson:2005vp}, if the entire TeV spectrum could be
attributed to hadronic processes, our methods yield an event rate of $N(E_\mu>
1$~TeV) $\sim 2.7$ yr$^{-1}$ in IceCube.  The Crab spectrum is relatively soft
and well described by leptonic processes~\cite{Aharonian:2004gb}, although
AMANDA observations hint at a neutrino signal~\cite{Ackermann:2005qg}.  A pionic
component may be uncovered by a significant observation of neutrino events.

The Vela complex that we have already discussed is located in the southern sky. 
Only after extensive observations was the existence of
Vela~Jr.~confirmed~\cite{Aschenbach}.  It would not be surprising, then, for a
similarly bright TeV source to be discovered in the northern sky (perhaps
directly through neutrino events).  The spectrum for such an object was
calculated near the end of Section~\ref{DetSection} for a variety of spectral
indices and cutoffs.  Future gamma-ray observations of this declination range in
the TeV regime may reveal another important source for neutrino astronomy.

Multi-wavelength observations of such a source would add a rung to a TeV
neutrino ``distance ladder".  Cosmic-ray interactions in the Earth's atmosphere
generate the nearest guaranteed neutrino source~\cite{Balkanov:1999up,
Ahrens:2002gq}.  Similar processes in the atmosphere of the Sun, which is at a
well known distance, are quite likely to produce TeV neutrinos as
well~\cite{Seckel91, Moskalenko:1991hm, Ingelman:1996mj}.  Of the prospective
sources, the Vela SNR is estimated to be only $\sim$~$250-300$~pc from Earth,
while Vela~Jr.~may be even closer~\cite{Cha, Dodson:2003ai, Aschenbach1999}.  RX
J1713--3946 has a distance estimate of $\sim 1$ kpc~\cite{Moriguchi:2005ty}.  At
$\sim 8$ kpc, the GCD is the most distant guaranteed source with a confirmed
distance.  AGN, if they produce measurable fluxes of neutrinos, would provide a
range of very remote point sources.  Measurements of neutrinos from a variety of
distances (ideally with flavor ratios) would provide important information for
testing neutrino properties at increasing $L/E$
(distance/energy)~\cite{Beacom:2002vi, Beacom:2003zg, Keranen:2003xd,
Beacom:2003eu, Hooper:2005jp, Hooper:2004xr, Christian:2004xb,
Anchordoqui:2005gj, Balaji:2006wi, Gupta:2004zx}.
%
%
\section{EGRET, the TeV Excess, and the Great Unknown}
The third EGRET catalog contains many unidentified gamma-ray sources detected in
the MeV--GeV range~\cite{Hartman:1999fc}.  In the Whipple survey of a selected
group of such EGRET sources, upper limits on high energy gamma-ray emissions
from a number of sources were determined~\cite{Fegan:2005rg}.  Two of these
sources, 3EG J1337+5029 and 3EG J2227+6122, possessed measured excesses that
were suggestive of gamma-ray emission~\cite{Fegan:2005ke}.  The upper limits
placed on these sources do not significantly constrain the EGRET data.  We
consider the possibility that these spectra are pionic and continue into the TeV
regime.  Spectral information, cutoffs, and muon rates are given in
Table~\ref{tab:egret}.  These would be readily detectable by IceCube.  We note
that 3EG J2227+6122 is roughly coincident with a position in the three-year
AMANDA point source survey which exhibits a modest excess of neutrino
events~\cite{Ackermann:2004ag}.  A less significant excess still appears in
newer data~\cite{Hulth:2006ny}.  In general, a maximum likelihood analysis
assuming the spectrum and normalization of such bright EGRET sources extended
into the TeV range may prove useful for neutrino searches.  Future observations
by GLAST, VERITAS, and MAGIC will provide valuable additional spectral 
information.

Recently, Milagro announced the detection of diffuse TeV emission from a $\sim
600$ deg$^2$ patch of the Galactic Plane that contains ten unidentified EGRET
point sources~\cite{Atkins:2005wu}.  This measurement places high energy
constraints on the spectra of these ten sources, as discussed in
Ref.~\cite{Prodanovic:2006bq}.  Consequently, these spectra, many of which are
quite hard, cannot all simply extrapolate to TeV energies.  HEGRA discovered an
unidentified TeV source in this region, however, it can account for only a small
fraction of the Milagro flux~\cite{Aharonian:2005ex}.  Located in the northern
sky, this region is accessible to observations by IceCube.  If this emission is
pionic and truly diffuse, then it would be very difficult to detect through
neutrino-induced muons~\cite{Kelley:2005py}.  Assuming that the entire flux is
instead divided among several point sources, the expected muon rate can be
calculated.  A detailed Milagro study of the Cygnus region revealed a
significant TeV flux~\cite{Smith06}.  Assigning all of the Milagro $3.5$~TeV
flux to a single source and extrapolating with a typical $\Gamma=2.2$ gamma-ray
spectrum (out to $100$~TeV), we arrive at a limiting case for the possible
neutrino emission from the region.  For a pionic spectrum, the calculated
$\nu_\mu$-induced muon rate is $N(E_\mu>1$~TeV) $\sim 2.3$ yr$^{-1}$ in IceCube.
 Future measurements of this region by Milagro, GLAST, VERITAS, and MAGIC will
determine whether multiple sources contribute to this diffuse flux.
%
\begin{table}[t]
\caption{\textit{Two Unidentified EGRET Sources:} Integrated ($\nu_\mu +
\bar{\nu}_\mu$)-induced muon rates, assuming a pionic spectrum, for a given
$E_\mu$ threshold and neutrino spectrum exponential cutoff $E_\nu^{cut}$ (in
TeV).  For one year of IceCube operation.  The assumed differential photon flux
at $1$~TeV is given in terms of $10^{-12}$ TeV$^{-1}$ cm$^{-2}$ s$^{-1}$, with
spectral index $\Gamma$.}
\begin{ruledtabular}
	\begin{tabular}{lcccrr}
	 Source & $\phi_\gamma$ & $\Gamma$ & $E_\nu^{cut}$ & $N_\mu(>0.5$~TeV) &
$N_\mu(>1$~TeV) \\
 \hline
	 3EG J1337+5029 & 20 & 2.2 & 10 & 4.1 & 2.5 \\
	                &    &     & 50 & 6.9 & 4.7 \\
	 3EG J2227+6122 & 10 & 2.0 & 10 & 3.0 & 1.9 \\
	                &    &     & 50 & 5.5 & 4.0 \\
 \end{tabular}
\end{ruledtabular}\label{tab:egret}
\end{table}
%
%
\section{Discussion and Conclusions}
Upcoming neutrino telescopes will deliver the first direct evidence concerning
the production mechanism of TeV gamma-ray sources.  Following the pioneering
AMANDA~\cite{Andres:1999hm} and Baikal~\cite{Belolaptikov:1997ry} efforts, the
next generation IceCube~\cite{Ahrens:2002dv} and
Mediterranean~\cite{Katz:2006wv} km$^3$ detectors will reach the scale necessary
to examine the Galactic sources discussed here.  Hadronic mechanisms would be
confirmed through neutrino detection, while a significant absence of neutrino
events would imply leptonic processes.  They will be able to probe the expected
high energy spectral cutoffs that would otherwise be unobservable.  It has been
suggested that the cutoff in a SNR may evolve with age~\cite{Volk:2004vi}. 
Finding these cutoffs with neutrinos may also yield valuable new information in
this regard.

Combined, IceCube and a km$^3$ Mediterranean detector will provide continuous,
all-sky coverage and can be used together to detect neutrino-induced muons and
showers from TeV sources.  The good angular resolution for muon events can be
used to precisely locate a source in neutrinos.  Showers, and contained muon
events, provide accurate reconstruction of the source spectrum at energies
beyond the reach of gamma-ray telescopes.  These measurements will provide
important information concerning the origin of high energy Galactic cosmic rays,
potentially directly observing the source population responsible for production
up to the knee at $\sim 3 \times 10^{15}$~eV.

The shell-type SNR Vela~Jr.~(RX J0852.0--4622) is an intriguing prospective
source of TeV neutrinos.  The extent and intensity of the TeV emission make
this, as well as the Vela complex as a whole, a unique target for neutrino
telescopes.  The confirmation of hadronic/leptonic processes through
detection/non-detection of neutrino events from Vela~X and the possible
detection of neutrinos through shower events from the Vela SNR shell are
exciting possibilities.

While it is expected that other sources may eventually be more compelling, only
the Galactic Center diffuse emission can presently be claimed to possess a
guaranteed neutrino flux.  The detection of this neutrino flux will give further
insight into the complex processes occurring in the Galactic Center region. 
When we consider that, to date, \textit{no high energy astrophysical neutrinos
have ever been positively detected}, the importance of such a guaranteed flux of
TeV neutrinos is difficult to overstate.

Included in the catalog of TeV sources are several that remain unidentified.  As
more observations are undertaken by the HESS, MAGIC, and VERITAS telescopes,
this number is expected to increase.  Discovering neutrino fluxes from these
sources would provide invaluable information concerning their nature.  As new
TeV gamma-ray sources are discovered, the pool of potential TeV neutrino sources
increases.  While some of these may not be significant alone, when grouped into
classes, stacking may increase the potential for neutrino studies.  An even
larger number of unidentified sources remains in the GeV regime from EGRET
observations.  Many of these sources have intense, hard spectra.  If any of
these sources, some of which are visible to IceCube, have spectra that
extrapolate into the TeV regime and are pionic, they would have abundant fluxes
of neutrinos.

In summary, the prospects for the near-term first discoveries of Galactic TeV
neutrino sources are very good.  Importantly, this conclusion is empirically
based on the measured spectra of bright Galactic TeV gamma-ray sources.  For
some of these, there are very compelling independent arguments that the observed
gamma rays arise from neutral pion decays, meaning that they must be accompanied
by neutrinos.  It is essential to test this directly for these sources, as well
as for others where the possibility of neutrino emission is uncertain.  We
emphasize the importance of using the measured muon energy spectra to
discriminate against the quickly falling atmospheric neutrino backgrounds.  For
example, a single event near $10$~TeV from a source direction is almost
certainly signal, while an event near $1$~TeV has a much higher probability of
being background.  This fact alone could be enough to help establish discovery. 
Due to the amplifying factors of neutrino cross section and muon range, neutrino
detectors have better reach to the highest energies in the source spectra, as
compared to gamma-ray telescopes, which can make precise measurements at lower
energies.  This complementarity can be exploited to help solve the long-standing
puzzle of the origin of the Galactic cosmic rays.

For all of the sources discussed here, the rates in km$^3$ neutrino telescopes
are relatively small, at most $\sim 1-10$~events/year, though we have argued
that even these small rates could have a powerful
impact.  While it remains possible that various uncertainties and detector
limitations may make these observations even more challenging, they might also
be better than described here.  If the source spectra extrapolate to higher
energies than we have assumed, or especially if the emitted TeV gamma-ray
spectra have been diminished by absorption in the sources, which is quite
possible, then the neutrino detection event rates could be significantly larger
than shown here.  Given the potentially unique power of neutrino astronomy, we
can only hope that Nature has been so kind. 

\begin{acknowledgments}
We are grateful to Felix Aharonian, Alexander Kappes, Paolo Lipari, Francesco
Vissani, Casey Watson and Hasan Y\"{u}ksel for helpful comments and discussions.
 MDK was supported by a Fowler Fellowship from The Ohio State University
Department of Physics.  JFB was supported by the National Science Foundation
under CAREER Grant No.~PHY-0547102, and by The Ohio State University.
\end{acknowledgments}

\newpage

\end{document}